\documentclass[onecolumn,amsmath,showpacs,11pt]{revtex4}
\usepackage{graphicx}
\usepackage{axodraw}
\usepackage{dcolumn}
\usepackage{bm}
\begin{document}
\newcommand{\hs}{\hspace*{0.5cm}}
\newcommand{\vs}{\vspace*{0.5cm}}
\newcommand{\be}{\begin{equation}}
\newcommand{\ee}{\end{equation}}
\newcommand{\bea}{\begin{eqnarray}}
\newcommand{\eea}{\end{eqnarray}}
\newcommand{\ben}{\begin{enumerate}}
\newcommand{\een}{\end{enumerate}}
\newcommand{\bde}{\begin{widetext}}
\newcommand{\ede}{\end{widetext}}
\newcommand{\nn}{\nonumber}
\newcommand{\crn}{\nonumber \\}
\newcommand{\Tr}{\mathrm{Tr}}
\newcommand{\non}{\nonumber}
\newcommand{\noi}{\noindent}
\newcommand{\al}{\alpha}
\newcommand{\la}{\lambda}
\newcommand{\bet}{\beta}
\newcommand{\ga}{\gamma}
\newcommand{\va}{\varphi}
\newcommand{\om}{\omega}
\newcommand{\pa}{\partial}
\newcommand{\+}{\dagger}
\newcommand{\fr}{\frac}
\newcommand{\bc}{\begin{center}}
\newcommand{\ec}{\end{center}}
\newcommand{\Ga}{\Gamma}
\newcommand{\de}{\delta}
\newcommand{\De}{\Delta}
\newcommand{\ep}{\epsilon}
\newcommand{\varep}{\varepsilon}
\newcommand{\ka}{\kappa}
\newcommand{\La}{\Lambda}
\newcommand{\si}{\sigma}
\newcommand{\Si}{\Sigma}
\newcommand{\ta}{\tau}
\newcommand{\up}{\upsilon}
\newcommand{\Up}{\Upsilon}
\newcommand{\ze}{\zeta}
\newcommand{\ps}{\psi}
\newcommand{\Ps}{\Psi}
\newcommand{\ph}{\phi}
\newcommand{\vph}{\varphi}
\newcommand{\Ph}{\Phi}
\newcommand{\Om}{\Omega}

\title{Left-right asymmetry and 750 GeV diphoton excess} 

\author{D. T. Huong}\email{dthuong@iop.vast.ac.vn}\affiliation{Institute of Physics, Vietnam Academy of Science and Technology, 10 Dao Tan, Ba Dinh, Hanoi, Vietnam}

\author{P. V. Dong}
\email {pvdong@iop.vast.ac.vn} \affiliation{Institute of Physics, Vietnam Academy of Science and Technology, 10 Dao Tan, Ba Dinh, Hanoi, Vietnam}

\begin{abstract}
We propose the left-right models based on $SU(3)_C\otimes SU(M)_L \otimes SU(N)_R \otimes U(1)_X$ gauge symmetry for $(M,N)=(3,3),\ (2,3)$, and (3,2) that address the 750 GeV diphoton excess recently reported by the LHC. The fermion contents are minimally introduced, and the generation number must match the fundamental color number to cancel anomalies and ensure QCD asymptotic freedom. The diphoton excess arises from the field that breaks the left-right symmetry spontaneously in the first model, while for the last models it emerges as an explicit violation of the left-right symmetry. The neutrino masses, flavor-changing neutral currents, and new boson searches are also discussed.      
\end{abstract}

\pacs{12.60.-i}
\date{\today}

\maketitle

\section{\label{intro}Introduction}

The minimal left-right model \cite{MLR,NLR} is defined by the gauge symmetry $SU(3)_C\otimes SU(2)_L\otimes SU(2)_R\otimes U(1)_{B-L}$ which preserves $SU(2)_L$ and $SU(2)_R$ interchange, called left-right symmetry. Consequently, $SU(2)_L$ and $SU(2)_R$ coupling constants equal, and every left-handed fermion doublet of $SU(2)_L$ corresponds to a right-handed fermion doublet of $SU(2)_R$ that composes those of the standard model. All such fermions have the color $(SU(3)_C)$ and baryon-minus-lepton $(U(1)_{B-L})$ quantum numbers, as usual. The minimal left-right model often works with a scalar bidoublet and two (a left and a right) scalar triplets. An interesting consequence is that the parity is exact, but its asymmetry as seen in the standard model is due to the spontaneous left-right symmetry breaking. Additionally, the seesaw mechanisms that generate small masses for the neutrinos naturally emerge since the right-handed neutrinos supplied as fundamental fermion constituents and the seesaw scales designated as the gauge symmetry breaking scales.    

The leading postulate \cite{diphotontheor} that solves the 750 GeV diphoton excess \cite{diphotonexp} recently observed at the LHC interprets the excess as a new neutral scalar that couples to extra heavy quarks (and/or new charged leptons/bosons in some cases). We will show that such a mechanism works in connection to fundamental left-right asymmetries. The minimal left-right model does not contain extra quarks as well as being unnaturally to explain the excess, so it should be extended. The first setup is based on $SU(3)_C\otimes SU(3)_L\otimes SU(3)_R\otimes U(1)_X$ gauge symmetry. The extra quarks exist as the third components of $SU(3)_L$ and $SU(3)_R$ triplets/antitriplets to complete the fermion representations, where the first two components are the standard model quarks. The new neutral scalar responsible for the 750 GeV diphoton excess becomes 33 component of a second scalar bitriplet that reflects a left-right symmetry from the first scalar bitriplet as including the usual scalar bidoublet. The second and third models are proposed, based on $SU(3)_C\otimes SU(2)_L\otimes SU(3)_R\otimes U(1)_X$ and $SU(3)_C\otimes SU(3)_L\otimes SU(2)_R\otimes U(1)_X$ gauge symmetries respectively, which explicitly violate the left-right symmetry. The nature of the scalar excess and heavy quarks radically differs from the first model. All the models require three generations of fermions in order to cancel anomalies as well as ensuring the QCD asymptotic freedom, analogous to the 3-3-1 model \cite{331m,331r}. Our setups have advantages and distinguishable phenomenologies, especially in the neutrino masses and flavor-changing neutral currents, compared to the known minimal left-right and 3-3-1 models. Our theories provide some new gauge bosons that possibly explain the recently-observed diboson anomalies \cite{dibosonexp} too.                            

In the next three sections, \ref{3l3r}, \ref{2l3r}, and \ref{3l2r}, we propose the models and correspondingly identify the mechanisms that explain the 750 GeV diphoton excess, respectively. The neutrino masses in each model are also discussed. Section \ref{pheno} is devoted to the flavor-changing neutral currents, new gauge-boson searches, and other aspects which improve the previously-studied theories. We also make conclusions and outlook in this section.     

\section{\label{3l3r} $SU(3)_C \otimes SU(3)_L \otimes SU(3)_R \otimes U(1)_X$ model}

The minimal left-right model cannot explain the 750 GeV diphoton excess. Let us suppose a higher gauge symmetry, $SU(3)_C \otimes SU(3)_L \otimes SU(3)_R \otimes U(1)_X$, completed by a $Z_2$ symmetry of $SU(3)_L$ and $SU(3)_R$ interchange, so-called left-right symmetry \footnote{Such a gauge symmetry and a fermion content with $q=0$ have been considered in \cite{dps}, but the scalar content to be studied is different.}. The fermion content which is anomaly free and reflects the $Z_2$ symmetry is given by
  \be \Psi_{aL} =
\left(\begin{array}{c}
               \nu_{aL}\\ e_{aL}\\ E^q_{aL}
\end{array}\right) \sim \left(1,3, 1,\frac{q-1}{3}\right),  \hs \hs  \Psi_{aR} =
\left(\begin{array}{c}
               \nu_{aR}\\ e_{aR}\\ E^q_{aR}
\end{array}\right) \sim \left(1,1, 3,\frac{q-1}{3}\right),\ee
\be Q_{\al L}=\left(\begin{array}{c}
  d_{\al L}\\  -u_{\al L}\\  J^{-q-\frac{1}{3}}_{\al L}
\end{array}\right)\sim \left(3,3^*,1,-\frac{q}{3}\right), \hs \hs
Q_{\al R}=\left(\begin{array}{c}
  d_{\al R}\\  -u_{\al R}\\  J^{-q-\frac{1}{3}}_{\al R}
\end{array}\right)\sim \left(3,1,3^*,-\frac{q}{3}\right), \ee   \be  Q_{3L}= \left(\begin{array}{c} u_{3L}\\  d_{3L}\\ J^{q+\frac{2}{3}}_{3L} \end{array}\right)\sim
 \left(3,3,1,\frac{q+1}{3}\right), \hs \hs  Q_{3R}= \left(\begin{array}{c} u_{3R}\\  d_{3R}\\ J^{q+\frac{2}{3}}_{3R} \end{array}\right)\sim
 \left(3,1,3,\frac{q+1}{3}\right).\ee
Here, $a=1,2,3$ and $\al=1,2$ are family indices. The
quantum numbers in the parentheses are given upon $(SU(3)_C,\ SU(3)_L,\ SU(3)_R,\ U(1)_X)$ groups, respectively. The extra fields $E_a$ and $J_a$ are new leptons and new quarks, respectively, whose electric charges depend on a free parameter~$q$. The electric charge operator also reflects the left-right symmetry, which is given by
\bea
Q = T_{3L}+T_{3R}+ \beta (T_{8L}+T_{8R})+X,
\eea
where $T_{iL,R}$ ($i=1,2,3,...,8$) and $X$ are $SU(3)_{L,R}$ and $U(1)_X$ charges respectively, and $\beta = \frac{-1-2q}{\sqrt{3}}$. Further, the $SU(3)_C$ charges will be denoted by $t_i$. Note that $X$ is not identical to $B-L$, which is unlike the minimal left-right model.

Provided that the left-handed fermion doublets are enlarged to become fundamental (triplet or antitriplet) representations of $SU(3)_L$ (note that the fermion representations of $SU(3)_R$ can be derived from the left-handed ones by a $Z_2$ transformation aforementioned), the $SU(3)_L$ anomaly cancelation requires the number of triplets equal to that of antitriplets, or in other words, the number of generations is a multiple of three---the number of fundamental colors. Since the extra quarks are included to complete the fundamental fermion representations, the number of generations is less than or equal to five to ensure the QCD asymptotic freedom. Consequently, the generation number is three, coinciding with the observation \cite{pdg}. It is easily to check that all other anomalies vanish. Such result is similar to the case of the 3-3-1 models \cite{331m,331r}. Our choice of the fermion representations is different from the ordinary trinification \cite{trinification}.                  

To break the gauge symmetries and generate the masses properly, we introduce two (a left and a right) scalar sextets,
 \bea \Sigma_L= \left(%
\begin{array}{ccc}
 \Sigma_{L 11}^0 & \frac{\Sigma_{L 12}^-}{\sqrt{2}}& \frac{\Sigma_{L13}^{q}}{\sqrt{2}} \\
  \frac{\Sigma_{L12}^-}{\sqrt{2}} & \Sigma_{L 22}^{--} & \frac{\Sigma_{L 23}^{q-1}}{\sqrt{2}} \\
 \frac{ \Sigma_{L 13}^{q}}{\sqrt{2}}& \frac{\Sigma_{L 23}^{q-1}}{\sqrt{2}}& \Sigma_{L 33}^{2q} \\
\end{array}%
\right)     = \left(1,6,1,\frac{2(q-1)}{3}\right), \eea \bea \Sigma_R= \left(%
\begin{array}{ccc}
 \Sigma_{R 11}^0 & \frac{\Sigma_{R 12} ^-}{\sqrt{2}}& \frac{\Sigma_{ R 13}^{q}}{\sqrt{2}} \\
  \frac{\Sigma_{R 12}^-}{\sqrt{2}} & \Sigma_{R 22}^{--} &\frac{ \Sigma_{R 23}^{q-1}}{\sqrt{2}} \\
 \frac{ \Sigma_{R 13}^{q}}{\sqrt{2}}&\frac{ \Sigma_{R 23}^{q-1}}{\sqrt{2}}& \Sigma_{R 33}^{2q} \\
\end{array}%
\right)     = \left(1,1,6,\frac{2(q-1)}{3}\right), \eea
and two scalar bitriplets,
\bea
\Phi_+= \left(%
\begin{array}{ccc}
 \Phi_{+ 11}^0 & \Phi_{+ 12}^+& \Phi_{+13}^{-q} \\
  \Phi_{+21}^- & \Phi_{+ 22}^{0} & \Phi_{+ 23}^{-q-1} \\
  \Phi_{+ 31}^{q}& \Phi_{+ 32}^{q+1}& \Phi_{+ 33}^{0} \\
\end{array}%
\right)     = (1,3,3^*,0), \eea
\bea
\Phi_-= \left(%
\begin{array}{ccc}
 \Phi_{- 11}^0 & \Phi_{- 12}^+& \Phi_{-13}^{-q} \\
 \Phi_{-21}^- & \Phi_{- 22}^{0} & \Phi_{- 23}^{-q-1} \\
  \Phi_{- 31}^{q}& \Phi_{- 32}^{q+1}& \Phi_{- 33}^{0} \\
\end{array}%
\right)     = (1,3^*,3,0). \eea Note that under the $Z_2$ symmetry, the scalars transform as $\Sigma_L\leftrightarrow \Sigma_R$ and $\Phi_+\leftrightarrow \Phi_-$, while under the gauge symmetry $SU(3)_L\otimes SU(3)_R$ they transform as $\Sigma_L\rightarrow U_L \Sigma_L U^T_L$, $\Sigma_R\rightarrow U_R \Sigma_R U^T_R$, $\Phi_+\rightarrow U_L\Phi_+ U^\dagger_R$, and $\Phi_- \rightarrow U_R\Phi_- U^\dagger_L$.  

The total Lagrangian of the considered model is given as
 \bea \mathcal{L} =\mathcal{L}_{\mathrm{kinetic}} +
\mathcal{L}_{\mathrm{Yukawa}}-V_{\mathrm{scalar}},\label{Ltotal1}\eea
where the first part provides kinetic terms and gauge interactions. The Yukawa terms are \bea \mathcal{L}_{\mathrm{Yukawa}} & =& x_{ab}( \bar{\Psi}^c_{aL} \Sigma^\dagger_L
\Psi_{bL}+ \bar{\Psi}^c_{aR} \Sigma^\dagger_R \Psi_{bR})+y_{+ab} \bar{\Psi}_{aL} \Phi_+ \Psi_{b R}
+y_{-ab} \bar{\Psi}_{aL} \Phi^\dagger_- \Psi_{bR}\crn
&&+z_{+33}\bar{Q}_{3L} \Phi_+ Q_{3R}+
z_{-33}\bar{Q}_{3L} \Phi^\dagger _- Q_{3R} +z_{+\al \beta}\bar{Q}_{\al L} \Phi^*_+ Q_{\beta
R}+z_{-\al \beta}\bar{Q}_{\al L} \Phi^T_- Q_{\beta R}\crn
&&+H.c. \eea The scalar potential that is invariant under the gauge and left-right symmetry as well as renormalizable can be divided into $V= V_\Sigma +V_\Phi +V_{\mathrm{mix}}$, where the first and second parts include only $(\Sigma_L,\Sigma_R)$ and $(\Phi_+,\Phi_-)$ respectively, while the last one contains the mixtures of $(\Sigma, \Phi)$, such as   \bea
 V_\Sigma & =& \mu_{\Sigma}^2\Tr(\Sigma_L\Sigma^\dag_L+\Sigma_R\Sigma^\dag_R)+\la_1 [(\Tr(\Sigma_L
 \Sigma^\dag_L))^2 +(\Tr(\Sigma_R
 \Sigma^\dag_R))^2]\crn &&+
  \la_2[\Tr(\Sigma_L\Sigma_L^\dag\Sigma_L\Sigma_L^\dag) +\Tr(\Sigma_R\Sigma_R^\dag\Sigma_R\Sigma_R^\dag)]
 + \la_3 \Tr(\Sigma_L \Sigma_L^\dag)\Tr(\Sigma_R \Sigma_R^\dag), \eea
 \bea
V_\Phi &=& \mu^2_1 \Tr( \Phi_+ \Phi_+^\dag+\Phi_- \Phi_-^\dag )+\mu_2^2 [\Tr(\Phi_+
\Phi_-)+H.c.]\crn
&&+\rho_1 [(\Tr(\Phi_+\Phi_+^\dag))^2+(\Tr(\Phi_-\Phi_-^\dag))^2]+ \rho_2 [(\Tr(\Phi_+
\Phi_-))^2+H.c.]
\crn
&&+ \rho_3 [\Tr(\Phi_+ \Phi_-) \Tr( \Phi_+
\Phi_+^\dag+\Phi_- \Phi_-^\dag )+H.c.]+\rho_4 \Tr( \Phi_+ \Phi_+^\dag)\Tr( \Phi_-
\Phi_-^\dag)
\crn
&&+\rho_5[\Tr(\Phi_+ \Phi_+^\dag\Phi_+\Phi_+^\dag)+ \Tr(\Phi_- \Phi_-^\dag\Phi_-\Phi_-^\dag)
]+\rho_6[ \Tr(\Phi_+ \Phi_- \Phi_+\Phi_- )+H.c.],
 \eea
\bea V_{\mathrm{mix}}&=&\kappa_1[
\Tr(\Phi_+\Phi_+^\dag)\Tr(\Sigma_L\Sigma_L^\dag)+\Tr(\Phi_-\Phi_-^\dag)\Tr(\Sigma_R\Sigma_R^\dag)]
\crn
&&+\kappa_2[\Tr(\Phi_+\Phi_+^\dag)\Tr(\Sigma_R\Sigma_R^\dag)+\Tr(\Phi_-\Phi_-^\dag)\Tr(\Sigma_L\Sigma_L^\dag )] \crn &&+
\kappa_3 [\Tr(\Phi_+\Phi_-)\Tr(\Sigma_L\Sigma_L^\dag+ \Sigma_R\Sigma_R^\dag
)+H.c.]\crn
&&+\kappa_4[\Tr(\Phi_+\Phi^\dagger_+ \Sigma_L \Sigma_L^\dagger)+\Tr(\Phi_-\Phi^\dagger_-\Sigma_R \Sigma^\dagger_R)] \crn &&+\kappa_5[\Tr(\Phi_+\Phi_- \Sigma_L \Sigma_L^\dagger)+\Tr(\Phi_-\Phi_+\Sigma_R \Sigma^\dagger_R)+H.c.]\crn
&&+ \kappa_6[\Tr(\Phi_+ \Sigma_R \Phi^*_- \Sigma_L^* )+H.c.].\eea 
 
The scalar potential is generally minimized at $\langle \Sigma_R\rangle =\fr{1}{\sqrt{2}}\mathrm{diag}(\La,0,0)$, $\langle \Sigma_L\rangle =\fr{1}{\sqrt{2}}\mathrm{diag}(\La',0,0)$, $\langle \Phi_+\rangle =\fr{1}{\sqrt{2}}\mathrm{diag}(u,v,w)$, and $\langle \Phi_-\rangle =\fr{1}{\sqrt{2}}\mathrm{diag}(u',v',w')$, where only the neutral fields can develop VEVs due to the $U(1)_Q$ invariance. We can choose the potential parameters so that $\La', u',v',w'\simeq 0$, while $\La, u,v,w\neq 0$, which break the left-right symmetry spontaneously, as desirable. The VEVs $(\La,w)$ break the gauge symmetry down to that of the standard model and give masses for the new particles. Subsequently, the VEVs $(u,v)$ break the standard model gauge symmetry down to $SU(3)_C\otimes U(1)_Q$ and provide the masses for ordinary particles. To be consistent with the standard model, we must suppose $u,v\ll w,\La$.     

The real or imaginary part of the neutral scalar $\Phi^0_{-33}=\fr{1}{\sqrt{2}}(S+i A)$ possibly explain the 750 GeV diphoton excess. Let us choose $S$, which couples to new fermions as 
\be \fr{1}{\sqrt{2}}\left(y_{-ab} \bar{E}_a E_b+z_{-ab}\bar{J}_aJ_b\right)S. \ee Notice that $E$ and $J$ get masses as $(m_E)_{ab}=-y_{+ab}\fr{w}{\sqrt{2}}$ and $(m_J)_{ab}=-z_{+ab}\fr{w}{\sqrt{2}}$. Due to the left-right symmetry, we have $h_+=h^\dagger_-$ $(h=y,z)$. The $S$ scalar is dominantly produced due to gluon fusion by $J$ loops at the LHC that is given at the leading order as  
\be \sigma(pp\rightarrow S)=\fr{\al^2_s m^2_S}{8 \pi^3 s }C_{gg}\left(\fr{1}{w}\right)^2,\ee which is independent of the Yukawa couplings $z_{\pm}$ due to the left-right symmetry. The dimensionless partonic integral was evaluated as $C_{gg}=2137$ at $\sqrt{s}=13$ TeV for $m_S=750$ GeV \cite{strumia}. Taking $\al_s=0.12$ and the QCD factor $K\simeq 2$, it follows \be \sigma(pp\rightarrow S)\simeq 0.32\times \left(\fr{1\ \mathrm{TeV}}{w}\right)^2\ \mathrm{pb}.\ee    

\begin{figure}[!h]
\begin{center}
\includegraphics[scale=0.7]{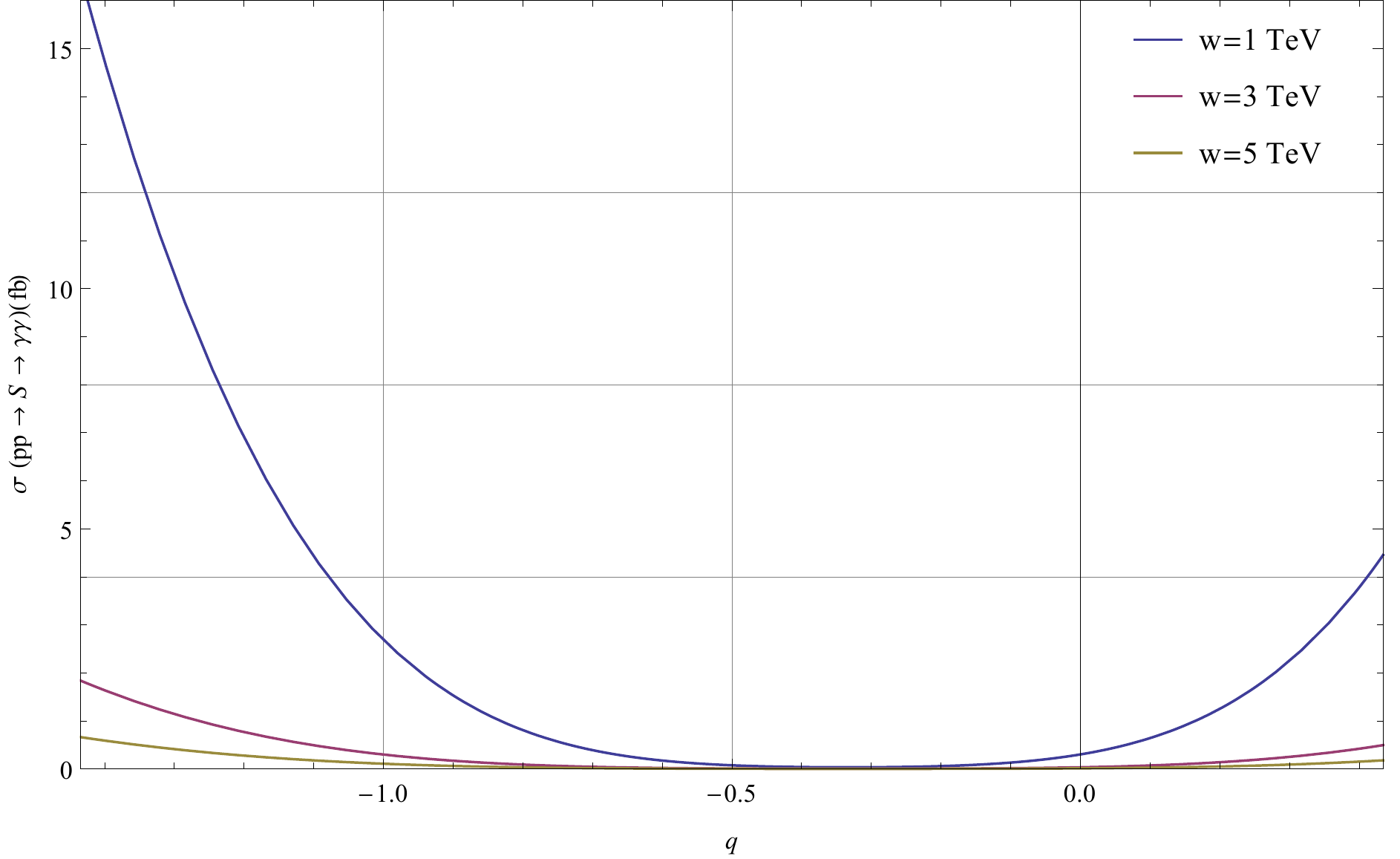}
\caption{\label{fig1}The cross-section for $pp\rightarrow S\rightarrow \gamma\gamma$ at the LHC $\sqrt{s}=13$ TeV and $m_S=750$ GeV.}
\end{center}
\end{figure}

The $S$ scalar mainly decays into two gluons as induced by $J$ loops, while the $\gamma\gamma$ mode that is induced by both $J$ and $E$ loops is smaller than. We compare 
\be \fr{\Ga(S\rightarrow \gamma\gamma)}{\Ga(S\rightarrow gg)}\simeq \fr 1 2 \left(\fr{\al}{\al_s}\right)^2\left|\sum_J Q^2_J+\fr 1 3 \sum_E Q^2_E\right|^2=\fr{2(6q^2+4q+1)^2}{9}\left(\fr{\al}{\al_s}\right)^2.\ee The photon field as renormalized, $\fr{1}{e^2}=\fr{1}{g^2_L}+\fr{1}{g^2_R}+\fr{\beta^2}{g^2_L}+\fr{\beta^2}{g^2_R}+\fr{1}{g^2_X}$, yields $\beta^2<-1+1/2s^2_W$, where $g_L=g_R=g$ and $s_W=e/g$. For $s^2_W=0.231$, it follows $-1.4345<q<0.4345$. We plot $\sigma(pp\rightarrow S\rightarrow \gamma\gamma)=\sigma(pp\rightarrow S)Br(S\rightarrow \gamma\gamma)$ as a function of $q$ for three values $w=1,3,5$ TeV as in Fig \ref{fig1}. We see that the cross-section is more enhanced when $w$ is small, in $\mathcal{O}(1)$ TeV, and $|q|$ is large, close to its bounds; and only in this case it can fit the data $\sigma(pp\rightarrow S\rightarrow \gamma\gamma)\sim 5 $ fb. This model predicts a narrow width of $S$ decay, $\fr{\Ga_S}{m_S}\simeq \fr{\al^2_s}{8\pi^3}(\fr{m_S}{w})^2\simeq 6\times 10^{-5}(\fr{m_S}{w})^2\ll 0.06$. The last value is favored by ATLAS. Further, the $w$ scale might be in tension with other bounds such as the FCNCs, dijet, and Drell-Yan processes at the LHC, which are briefly considered in Sec. \ref{pheno}.     

Another possibility is to interpret the real part of $\Phi^0_{+33}$ as the 750 GeV diphoton excess (note that its imaginary part is a Goldstone boson). The production cross-section of this candidate is the same previous case, however its branching decay ratio into two photons should be radically smaller than the case above since it has a VEV and thus mixes with the standard model Higgs boson, which significally decays into the the standard model heavy particles such as $t\bar{t}$, $hh$, $WW$, and $ZZ$. Without a fine-tuning in the scalar couplings to omit the mixing effects \cite{dn}, the total cross-section would be too small to fit the data in some fb for the $w$ scale in TeV. This candidate is disfavored, but a similar candidate is viable for the model below.            

\section{\label{2l3r} $SU(3)_C \otimes SU(2)_L \otimes SU(3)_R \otimes U(1)_X$ model}

The fermion content of the $SU(3)_C \otimes SU(2)_L \otimes SU(3)_R \otimes
 U(1)_X$ model that is anomaly free is given as follows
 \bea
\Psi_{aL}=\left(%
\begin{array}{c}
  \nu_{aL} \\
  e_{aL}\\
\end{array}%
\right)\sim \left(1,2,1,-\frac{1}{2}\right), \hs \hs \Psi_{aR}=\left(%
\begin{array}{c}
  \nu_{aR} \\
 e_{aR} \\
  E_{aR}^q \\
\end{array}%
\right)\sim \left(1,1,3,\frac{q-1}{3}\right),
 \eea
 \bea
Q_{3L}= \left(%
\begin{array}{c}
  u_{3L} \\
  d_{3L} \\
\end{array}%
\right) \sim \left(3,2,1,\frac{1}{6}\right), \hs \hs Q_{3R}= \left(%
\begin{array}{c}
  u_{3R} \\
  d_{3R} \\
  J^{q+\frac{2}{3}}_{3R} \\
\end{array}%
\right)\sim \left(3,1,3,\frac{q+1}{3}\right),
 \eea
 \bea
 Q_{\al L} =\left(%
\begin{array}{c}
  u_{\al L} \\
  d_{\al L} \\
\end{array}%
\right)\sim \left(3,2,1,\frac{1}{6}\right), \hs \hs Q_{\al R}=\left(%
\begin{array}{c}
  d_{\al R} \\
  -u_{\al R} \\
  J_{\al R}^{-q-\frac{1}{3}} \\
\end{array}%
\right)\sim \left(3,1,3^*,-\frac{q}{3}\right),
 \eea
 \bea
 E^q_{aL}\sim (1,1,1,q), \hs J^{q+\fr 2 3}_{3L}\sim \left(3,1,1,q+\fr{2}{3}\right), \hs J^{-q-\fr 1 3}_{\al L} \sim
 \left(3,1,1,-q-\fr{1}{3}\right),
 \eea
where $a=1,2,3$ and $\al=1,2$ are family indices, $E_a$ and $J_a$ are new fermions, and $q$ is an electric charge parameter, similarly to the previous model. The electric charge operator takes the form,
\bea
Q =T_{3L}+ T_{3R}+ \beta T_{8R}+X, \label{electric1}
\eea
where $T_{aL}$, $T_{iR}$ ($i=1,2,3,...,8$), and $X$ are $SU(2)_L$, $SU(3)_R$, and $U(1)_X$ charges, respectively, and $\beta=-(1+2q)/\sqrt{3}$ in similarity to the above model.  

We see that the gauge symmetry and fermion content contain those of the minimal left-right model, but the theory does not conserve the left-right symmetry. Therefore, the $SU(2)_L$ and $SU(3)_R$ gauge couplings are generally unrelated, and the electric charge $q$ (or $\beta$) is arbitrary, as the photon field is renormalized. Again, the number of fermion generations must be three as a consequence of $SU(3)_R$ anomaly cancelation and QCD asymptotic freedom.       

The gauge symmetry breaking and mass generation are properly done by using the following scalar multiplets, 
 \bea
S= \left(%
\begin{array}{ccc}
  S_{11}^0 & S_{12}^+ & S_{13}^{-q} \\
  S_{21}^- & S_{22}^0 & S_{23}^{-1-q} \\
\end{array}%
\right) \sim \left(1,2,3^*,-\frac{2q+1}{6}\right) , \hs \phi=\left(%
\begin{array}{c}
  \phi_1^{-q} \\
  \phi_2^{-q-1} \\
  \phi_3^0 \\
\end{array}%
\right)\sim \left(1,1,3,-\frac{1+2q}{3}\right),  \eea \bea \Xi= \left(%
\begin{array}{ccc}
  \Xi^0_{11} &\frac{ \Xi_{12}^-}{\sqrt{2}} &\frac{ \Xi_{13}^q }{\sqrt{2}}\\
  \frac{ \Xi_{12}^-}{\sqrt{2}} & \Xi_{22}^{--} & \frac{\Xi_{23}^{q-1}}{\sqrt{2}} \\
  \frac{\Xi_{13}^q}{\sqrt{2}} & \frac{\Xi_{23}^{q-1}}{\sqrt{2}} & \Xi_{33}^{2q}\\
\end{array}%
\right)\sim \left(1,1,6,\frac{2(q-1)}{3}\right), \hs \hs \Delta =\left(%
\begin{array}{cc}
  \Delta_{11}^0 & \frac{\Delta_{12}^-}{\sqrt{2}} \\
 \frac{\Delta_{12}^-}{\sqrt{2}} & \Delta_{22}^{--} \\
\end{array}%
\right)\sim (1,3,1,-1), \eea 
where the introduction of $\Delta$ implies a combinational seesaw mechanism of type I and II for the neutrino masses. Otherwise, if $\Delta$ is omitted, the neutrinos gain masses only from the type I seesaw mechanism. Moreover, $\Xi_{11}$ and $\phi_3$ when develop VEVs, $\langle \Xi_{11}\rangle =\fr{1}{\sqrt{2}}\La$ and $\langle \phi_{3}\rangle =\fr{1}{\sqrt{2}}w$, will break the gauge symmetry down to that of the standard model and provide masses for new particles such as $\nu_R$, $E$, $J$, and new gauge bosons. Whereas, the VEVs of $S_{11}$ and $S_{22}$, $\langle S_{11}\rangle =\fr{1}{\sqrt{2}}u$ and $\langle S_{22}\rangle =\fr{1}{\sqrt{2}}v$, break the standard model gauge symmetry down to $SU(3)_C\otimes U(1)_Q$ and generate masses for ordinary particles. For consistency with the standard model, we assume $u,v\ll w,\La$.

Apart from the kinetic part that including gauge interactions, the Yukawa Lagrangian and scalar potential are obtained as 
 \bea
 \mathcal{L}_{\mathrm{Yukawa}} & =&  h^l_{ab} \bar{\Psi}_{aL}S \Psi_{bR}+h^{L}_{ab} \bar{\Psi}^c_{aL} \Delta^\dagger \Psi_{bL} 
 +h^{R}_{ab}
\bar{\Psi}^c_{aR} \Xi^\dagger \Psi_{bR}\crn
&& + h_{a3}^q\bar{Q}_{aL}S Q_{3R}+h^q_{a \beta}\bar{\tilde{Q}}_{a L} S^* Q_{\beta R}
 + h_{ab}^E\bar{E}_{aL}\phi^\dagger \Psi_{bR}  \nonumber \\
&&+ h^J_{33}\bar{J}_{3L}\phi^\dagger Q_{3R} + h^J_{\al \beta} \bar{J}_{\al L} \phi^T Q_{\beta R}+H.c., \label{231}\eea  
\bea V &= &\mu_S^2 \Tr(S^\dag S)+\la_{1S}[\Tr (S^\dag
S)]^2+\la_{2S}\Tr(S^\dag SS^\dag S) +\mu_\Xi^2\Tr(\Xi^\dag \Xi)\crn
&& +\la_{1\Xi}[ \Tr(\Xi^\dag \Xi)]^2
+\la_{2 \Xi}\Tr( \Xi^\dag \Xi \Xi^\dag \Xi)+\mu_\Delta^2 \Tr(\Delta^\dag
\Delta)+\la_{1\Delta}\Tr(\Delta^\dag \Delta)^2\crn
&&+\la_{2 \Delta}\Tr(\Delta^\dag \Delta\Delta^\dag
\Delta) +\la_{3\Delta} \Tr{(\Delta^\dagger \Delta^\dagger)}\Tr{(\Delta\Delta)}+\mu_\phi^2 \phi^\dag \phi +\la_\phi (\phi^\dag \phi)^2\crn
&&+\la_{\phi S}
(\phi^\dag \phi) \Tr(S^\dag S)+\la_{\phi \Xi} (\phi^\dag \phi) \Tr(\Xi^\dag \Xi)+\la_{\phi \Delta}
(\phi^\dag \phi) \Tr(\Delta^\dag \Delta) \crn
&& + \la_{ \Xi S}\Tr(\Xi^\dag \Xi) \Tr(S^\dag
S)+\la_{ \Xi \Delta}\Tr(\Xi^\dag \Xi) \Tr(\Delta^\dag \Delta) + \la_{ \Delta S}\Tr(\Delta^\dag \Delta) \Tr(S^\dag S).  \label{232} \eea Above, we denoted $\tilde{Q}_L=i\sigma_2 Q_L$ that transforms as $2^*$ under $SU(2)_L$, $\tilde{Q}_{L}\rightarrow U^*_L \tilde{Q}_L$. Note also that $S\rightarrow U_LS U^\dagger_R$, $\Xi\rightarrow U_R \Xi U^T_R$, $\Delta\rightarrow U_L \Delta U^T_L$, and $Q_{\al R}\rightarrow U^*_R Q_{\al R}$, under $SU(2)_L\otimes SU(3)_R$.  

Let us interpret the 750 GeV diphoton excess. From the Yukawa interactions, there is generally no
mixing between the ordinary quarks and new quarks, and no mixing between the ordinary charged-leptons and new leptons. The third component of $\phi$ triplet is a standard model singlet which dominantly couples to the new quarks and to the new leptons. The imaginary part of $\phi_3$ is a Goldstone boson, but its real part is a physical Higgs boson to be identified as the 750 GeV diphoton excess, called $X$. Moreover, $X$ interacts with the $SU(3)_R$ gauge bosons as well as other new scalars. To keep the rate $X\rightarrow \gamma\gamma$ reasonably large, the $SU(3)_R$ breaking scales should be high so that $X$ cannot decay into the new particles as kinematically suppressed. This also ensures that the dangerous FCNCs, dijet and Drell-Yan processes are prevented.   

\begin{figure}[!h]
\begin{center}
\includegraphics[scale=0.8]{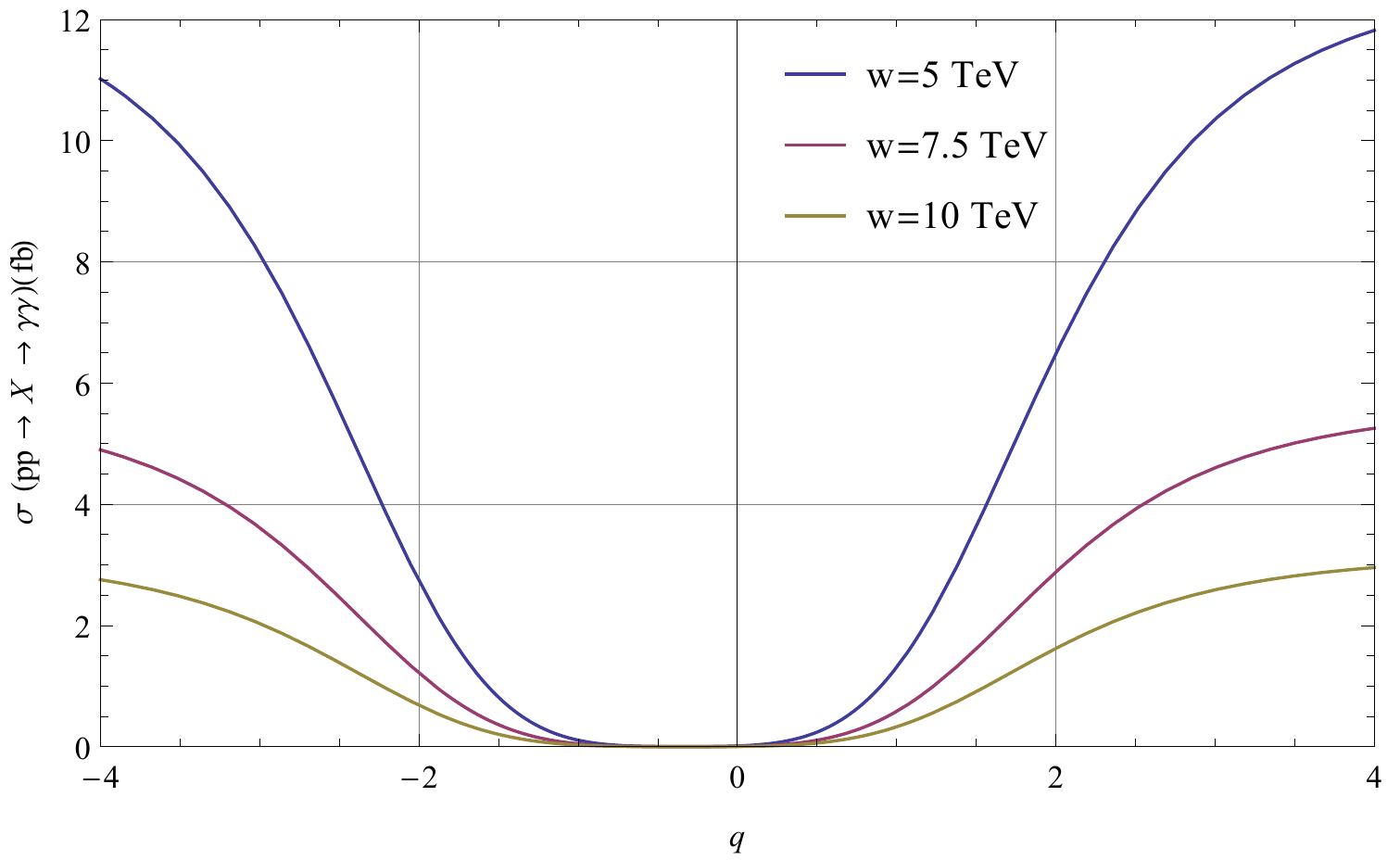}
\caption{\label{fig2}The cross-section for $pp\rightarrow X\rightarrow \gamma\gamma$ at the LHC $\sqrt{s}=13$ TeV and $m_X=750$ GeV.}
\end{center}
\end{figure}

Further, from the scalar potential there might exist a mixing between the Higgs singlet $X$ and the standard model Higgs boson. This mixing is expected to be small when its VEV $w$ is so large or a fine-tuning in scalar couplings is needed~\cite{dn}. The former leads to a too small cross-section $\sigma(pp\rightarrow X)$ to compare with the data, so it is not imposed. In the case under consideration, the couplings of $X$ to the standard model fermions, Higgs, and gauge bosons can be neglected. The $X$ excess decays only into gluons (via $J$ loops) and photons (via $J$ and $E$ loops), completely similar to the previous model. But, the difference is that $w$ is sizable and $q$ is arbitrary. The total cross-section $\sigma(pp\rightarrow X\rightarrow \gamma\gamma)$ is given in Fig. \ref{fig2}, which fits the data when $q$ is actually large. This case favors a narrow width of $X$ decay $\fr{\Ga_X}{m_X}\simeq 6\times 10^{-5}(\fr{m_X}{w})^2\ll 0.06$ as suggested by ATLAS. 

\begin{figure}[!h]
\begin{center}
\includegraphics[scale=0.8]{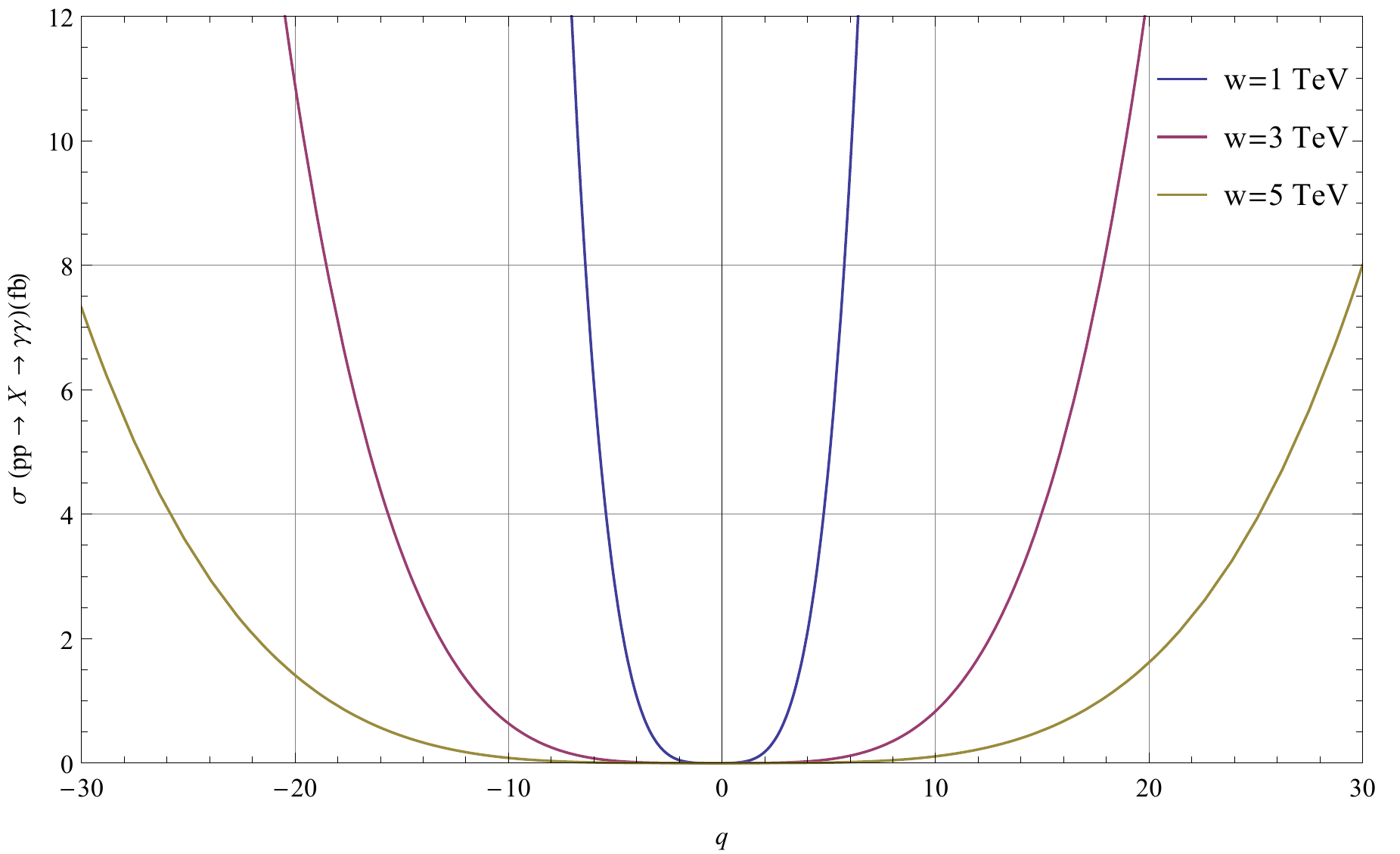}
\caption{\label{fig3}The cross-section for $pp\rightarrow X\rightarrow \gamma\gamma$ at the LHC $\sqrt{s}=13$ TeV and $m_X=750$ GeV.}
\end{center}
\end{figure}

When keeping the VEV $w$ in a few TeV and the scalar couplings are relaxed, the large decays of $X$ into $WW$, $ZZ$, $hh$, and $t\bar{t}$ dominating over the gluon mode are expected due to the mixing, and it can fit the large width as suggested by ATLAS \cite{dn}. Therefore, the branching decay ratio of $X$ into two photons is $Br(X\rightarrow \gamma\gamma)\simeq \Ga(X\rightarrow \gamma\gamma)/45\mathrm{GeV}$, where
\be \Ga(X\rightarrow \gamma\gamma)=\fr{\al^2 m^3_X}{16\pi^3 w^2}\left|\sum_J Q^2_J+\fr 1 3 \sum_E Q^2_E\right|^2.\ee
We plot the total cross-section as a function of $q$ for $w=1,3,5$ TeV in Fig. \ref{fig3}. Consequently, the signal strength fits the data if $|q|$ is large, corresponding to each $w$.        

\begin{figure}[!h]
\begin{center}
\includegraphics[scale=0.8]{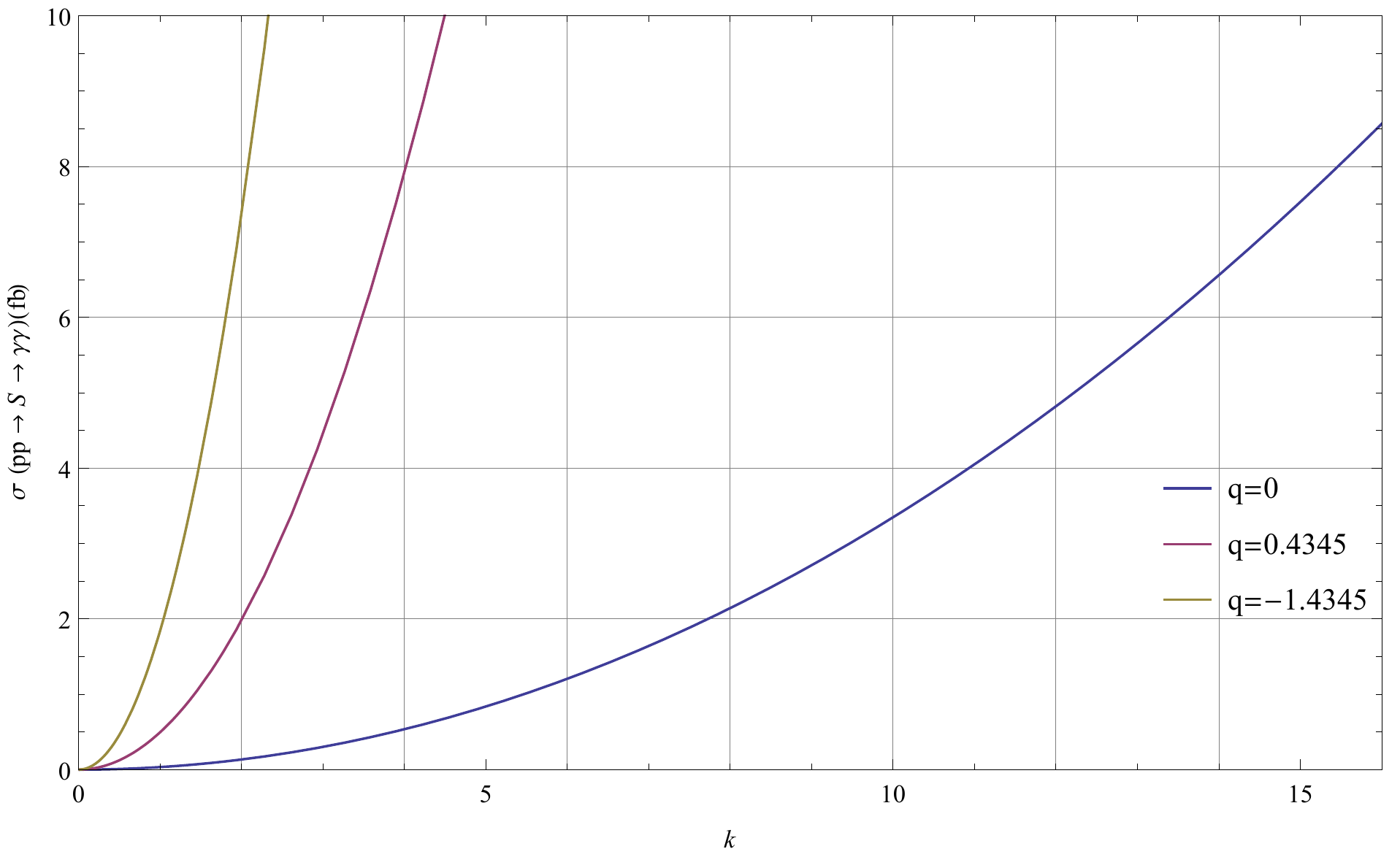}
\caption{\label{fig4}The cross-section for $pp\rightarrow S\rightarrow \gamma\gamma$ at the LHC $\sqrt{s}=13$ TeV and $m_X=750$ GeV. The $w$ scale is taken as $w=3$ TeV.}
\end{center}
\end{figure}

Some remarks are given in order,
\ben
\item The advantage of the considering model over the model in the previous section as well as the 3-3-1 models \cite{dn} is that  the parameter $q$ is not bounded. This relaxation of $q$ makes the candidate we interpreted as the 750 GeV diphoton excess viable. 
\item To avoid the unwanted low-values of $w$ and the disfavored large values of $q$, we can introduce an inert scalar triplet $\phi' \sim (1,1,3,-\frac{1+2q}{3})$ which has the gauge quantum numbers similarly to $\phi$. The field $\phi'$ couples to the leptons and quarks like $\phi$, but it does not develop VEV, which can be ensured by choosing appropriate potential parameters. At this stage, we can interpret the real or imaginary part of $\phi'_3$ as the 750 GeV diphoton excess instead of $X$. Taking the real part, called $S$, into account the signal strength is naively proportional to 
\be \sigma(pp\rightarrow S\rightarrow \gamma\gamma)\sim \left(\fr{h'^F}{h^F}\right)^2\sigma(pp\rightarrow X\rightarrow \gamma\gamma),\ee where $h'^F$ and $h^F$ ($F=E,J$) are the Yukawa couplings of $S$ and $X$ to the new quarks or new charged-leptons, respectively. Therefore, the total cross-section is more enhanced when $h'^F/h^F$ is large. Assuming that the Yukawa coupling ratios are equivalent for any new quarks and new charged-leptons, $k=h'^F/h^F$, we plot the total cross-section as a function of $k$ for $q=0$ (the model without exotic charges), $q=0.4345$ and $q=-1.4345$ (the bounds of the previous model) as in Fig. \ref{fig4}. Hence, when $q=0$, $k\sim 12$ is required to fit the data $\sigma_{\gamma\gamma}\sim 5$ fb, whereas when $q=-1.4345$ we only need $k\sim 1.5$ to have $\sigma_{\gamma\gamma}\sim 5$ fb. The scenario with the inert scalar $\phi'$ yields a narrow width for the 750 GeV diphoton excess.  
In summary, a large ratio between $h^{\prime F}$ and $h^F$ can explain the observed excess while still keeping the  $SU(3)_R$ symmetry breaking scale $w$ high enough to satisfy the other
constrains as well as the electric charge value of the new fermions not too large.
\item
In the $SU(3)_C \otimes SU(3)_L \otimes U(1)_X$ model, the electric charge operator takes the form, $Q=T_3+ \beta T_8 + X$, where $\beta = -\frac{1+2q}{\sqrt{3}}$, and the $q$ parameter as the electric charge of the third component of the left-handed lepton triplet is constrained by $-2.08\leq
q\leq1.08$ \cite{beta331}. Hence, if the scalar field that is used to break $SU(3)_L$ symmetry is interpreted as the 750 GeV diphoton excess, its status is quite similar to the second candidate in the model of the previous section~\cite{dn}. However, if one introduces an inert scalar field instead it may fit the 750 GeV diphoton excess, similarly to $\phi'$ in the second remark above.    
\een 

To conclude, the 750 GeV diphoton excess can be explained in the $SU(3)_C \otimes SU(2)_L \otimes SU(3)_R \otimes U(1)_X$
model of TeV new physics scale by appropriate large electric charges for the new fermions or alternatively introducing an inert scalar triplet instead. Referred to the higher symmetry of the previous model, $\Phi_+=S\oplus \phi^*$ while $\Phi_-\supset \phi'$, therefore $h^F\neq h'^F$ is a consequence of the explicit left-right asymmetry, in the same reason for $q$ as unconstrained. Both the candidates are thus interpreted as a result of the obvious left-right asymmetry.      

\section{\label{3l2r} $SU(3)_C \otimes SU(3)_L \otimes SU(2)_R \otimes U(1)_X$ model}
The apparent left-right asymmetry can also be achieved by introducing the following gauge symmetry $SU(3)_C \otimes SU(3)_L \otimes SU(2)_R \otimes U(1)_X$. In this case, the electric charge operator is
 \bea
 Q= T_{3L}+\beta T_{8L} +T_{3R} +X, \label{charge3}
 \eea
 where $T_{iL}$ ($i=1,2,3,...,8$) and $T_{aR}$ ($a=1,2,3$) are the generators of $SU(3)_L$ and $SU(2)_R$ groups, respectively, and $X$ is the generator of $U(1)_X$, as usual. 
 
 The fermion content which is anomaly free is given by
\bea
 \Psi_{aL}=\left(%
\begin{array}{c}
  \nu_{aL} \\
 e_{aL} \\
  E_{aL}^q \\
\end{array}%
\right)\sim \left(1,3,1,\frac{q-1}{3}\right), \hs \Psi_{aR}=\left(%
\begin{array}{c}
  \nu_{aR} \\
  e_{aR}\\
\end{array}%
\right)\sim \left(1,1,2,-\frac{1}{2}\right),
 \eea
 \bea
 Q_{3L}= \left(%
\begin{array}{c}
  u_{3L} \\
  d_{3L} \\
  J^{q+\frac{2}{3}}_{3L} \\
\end{array}%
\right)\sim \left(3,3,1,\frac{q+1}{3}\right), \hs Q_{3R}= \left(%
\begin{array}{c}
  u_{3R} \\
  d_{3R} \\
\end{array}%
\right) \sim \left(3,1,2,\frac{1}{6}\right),
 \eea
 \bea
 Q_{\al L}=\left(%
\begin{array}{c}
  d_{\al L} \\
  -u_{\al L} \\
  J_{\al L}^{-q-\frac{1}{3}} \\
\end{array}%
\right)\sim \left(3,3^*,1,-\frac{q}{3}\right), \hs Q_{\al R} =\left(%
\begin{array}{c}
  u_{\al R} \\
  d_{\al R} \\
\end{array}%
\right)\sim \left(3,1,2,\frac{1}{6}\right),
 \eea
 \bea
 E^q_{aR}\sim (1,1,1,q), \hs J^{q+\fr 2 3}_{3R}\sim \left(3,1,1,q+\fr{2}{3}\right), \hs J^{-q-\fr 1 3}_{\al R} \sim
 \left(3,1,1,-q-\fr{1}{3}\right),
 \eea
where $E$ and $J$ are new charged leptons and new quarks, and $\al=1,2$, as usual. The model requires three generations of fermions in order to validate the $SU(3)_L$ anomaly cancelation and QCD asymptotic freedom. The other anomalies are also canceled out.    

Quite analogous to the previous model, the scalar multiplets are properly imposed as follows
 \bea
S= \left(%
\begin{array}{ccc}
  S_{11}^{0} & S_{12}^{ +} & S_{13}^{ -q} \\
  S_{21}^{ -} & S_{22}^{ 0} & S_{23}^{ -1-q} \\
\end{array}%
\right) \sim \left(1,3,2^*,-\frac{2q+1}{6}\right) , \hs \phi =\left(%
\begin{array}{c}
  \phi_1^{ -q} \\
  \phi_2^{-q-1} \\
  \phi_3^{ 0}\\
\end{array}%
\right)\sim \left(1,3,1,-\frac{1+2q}{3}\right),  \eea \bea \Xi= \left(%
\begin{array}{ccc}
  \Xi^{ 0}_{11} &\frac{ \Xi_{12}^{ -}}{\sqrt{2}} &\frac{ \Xi_{13}^{ q} }{\sqrt{2}}\\
  \frac{ \Xi_{12}^{ -}}{\sqrt{2}} & \Xi_{22}^{ --} & \frac{\Xi_{23}^{ q-1}}{\sqrt{2}} \\
  \frac{\Xi_{13}^{ q}}{\sqrt{2}} & \frac{\Xi_{23}^{ q-1}}{\sqrt{2}} & \Xi_{33}^{ 2q}\\
\end{array}%
\right)\sim \left(1,6,1,\frac{2(q-1)}{3}\right), \hs \Delta =\left(%
\begin{array}{cc}
  \Delta_{11}^{ 0} & \frac{\Delta_{12}^{ -}}{\sqrt{2}} \\
 \frac{\Delta_{12}^{ -} }{\sqrt{2}} & \Delta_{22}^{ --} \\
\end{array}%
\right)\sim (1,1,3,-1). \eea Here, $\Xi$ can be included or not. Its presence implies the type II seesaw mechanism for neutrino masses. Otherwise, the neutrinos always get small masses via the type I seesaw due to $\Delta$.  

The Yukawa Lagrangian and scalar potential are respectively obtained as 
 \bea
 \mathcal{L}_{\mathrm{Yakawa}} & =&  h_{ab}^{l} \bar{\Psi}_{aL}S  \Psi_{bR}+h_{ab}^L \bar{\Psi}^c_{aL} \Xi^\dagger\Psi_{bL}+h_{ab}^R \bar{\Psi}^c_{aR} \Delta^\dagger \Psi_{bR} \crn
 &&+h^{q}_{\al a}\bar{Q}_{\al L} S^*  \tilde{Q}_{a R}
 + h_{3a}^{q}\bar{Q}_{3L}S Q_{aR}+ h_{ab}^{E}\bar{\Psi }_{a L}\phi E_{bR} \nonumber \\
&&+h^{J}_{33} \bar{Q}_{3L} \phi J_{3R}+ h^{J}_{\al \beta} \bar{Q}_{\al L}
\phi^* J_{\beta R} +H.c.,\eea 
\bea V &= &\mu_S^2 \Tr(S^{\dag}
S)+\la_{1S}[\Tr(S^{\dag} S)]^2+\la_{2S}\Tr(S^{\dag} S S^{\dag} S)+\mu_{\Xi}^{2}\Tr(\Xi^{\dag} \Xi)\crn
&& +\la_{1\Xi}
[\Tr(\Xi^{\dag} \Xi)]^2 +\la_{2 \Xi}\Tr(\Xi^{\dag}
\Xi \Xi^{\dag} \Xi)+\mu_{\Delta }^2 \Tr(\Delta^{\dag}
\Delta)+\la_{1\Delta}[\Tr(\Delta^\dag \Delta)]^2\crn
&& +\la_{2 \Delta
}\Tr(\Delta^{\dag} \Delta \Delta^{\dag} \Delta)+\la_{3\Delta}\Tr{\Delta^\dagger \Delta^\dagger}\Tr(\Delta\Delta) +\mu_\phi^2 \phi^{\dag} \phi  +\la_{\phi} (\phi^{\dag}
\phi)^2\crn
&&+\la_{\phi S} (\phi^{\dag} \phi) \Tr(S^{\dag
}S)+\la_{\phi \Xi}(\phi^{\dag} \phi) \Tr(\Xi^{\dag}
\Xi)+\la_{\phi\Delta}(\phi^{\dag} \phi) \Tr(\Delta^{
\dag} \Delta)  \nonumber \\ && + \la_{ \Xi S}\Tr(\Xi^{\dag} \Xi)
\Tr(S^{\dag} S)+\la_{ \Xi \Delta}\Tr(\Xi^{\dag} \Xi)
\Tr(\Delta^{\dag} \Delta)+\la_{\Delta S}\Tr(\Delta^{\dag}
\Delta) \Tr(S^{\dag} S). \nonumber  \eea
Above, noting that under $SU(3)_L\otimes SU(2)_R$ the fields transform as $S\rightarrow U_L S U^\dagger_R$, $\Xi\rightarrow U_L \Xi U^T_L$, $\Delta\rightarrow U_R \Delta U^T_R$, $Q_{\al L}\rightarrow U^*_L Q_{\al L}$, and $\tilde{Q}_R=i\sigma_2 Q_R\rightarrow U^*_R \tilde{Q}_R$. 

The photon field as renormalized yields a bound on $\beta$ such as $|\beta|<\cot_W$, thus $-2.08<q<1.08$ for $s^2_W=0.231$, analogous to the 3-3-1 model. The 750 GeV diphoton excess can be identified as the real part of $\phi_3^0$, called $X$, (its imaginary part is a Goldstone boson) or alternatively an inert scalar field as a copy of $\phi_3$, called $S$, which all have Yukawa couplings to the new fermions as usual. Referred to the cases of the 3-3-1 models \cite{dn}, all the conclusions drawn for the 750 GeV diphoton excess exactly apply for this model, as already mentioned in the previous section. That is being said, $X$ is a good candidate for the excess if $q$ is close to the bounds, and the new physics scale is low, in $\mathcal{O}(1)$ TeV, which must be in tension with other constraints from the FCNCs, dijet and Drell-Yan searches. Otherwise, such constraints can be evaded if $S$ is interpreted as the excess instead. But, in this case, a hierarchy in the Yukawa couplings of $S$ and $X$ to the new quarks and charged-leptons might be required, again as a consequence of the explicit left-right asymmetry.    

\section{\label{pheno} Other aspects and outlook} 

The three models proposed so far have the property that the third quark generation transforms differently from the first two, under the gauge symmetries. Hence, there are tree-level FCNCs associated with the new neutral gauge bosons, which non-universally couple to ordinary quarks. For the gauge symmetry, $SU(3)_C\otimes SU(3)_L\otimes SU(3)_R\otimes U(1)_X$, such new neutral gauge bosons are $A_{8L}$, $A_{8R}$, and $B$, as coupled to the non-universal charges, $T_{8L}$, $T_{8R}$, and $X$, respectively. Moreover, $X$ is related to $T_8$'s via the electric charge operator. Therefore, the FCNCs are associated with the fields, $Z'_{L,R}=(A_{8L,8R}-\beta t_X B)/\sqrt{1+\beta^2 t^2_X}$ with $t_X=g_X/g$, as obtained by  
\be \mathcal{L}_{\mathrm{FCNC}}=-\fr{g}{\sqrt{3}}\sqrt{1+\beta^2 t^2_X}\left[(V^*_{qL})_{3i}(V_{qL})_{3j} \bar{q}_{iL}\ga^\mu q_{jL}Z'_{L\mu}+(L\rightarrow R)\right],\ee where $i\neq j$, $q=(u,c,t)$ or $q=(d,s,b)$, and $V_{qL,qR}$ are left-handed, right-handed quark mixing matrices which relate the gauge states to mass eigenstates. For the second model above, the left part in the FCNCs is omitted, whereas for the third model the right part is omitted. The mixing effects of $Z'_{L,R}$ with the standard model $Z$ negligibly change the FCNCs. However, the contribution of $Z_R$ gauge boson to the FCNCs can be large due to the large mixings with $Z'_{L,R}$.

The above FCNCs contribute to neutral meson mixings as well as rare semileptonic/leptonic meson decays. Taking a strong bound coming from $B^0_s$-$\bar{B}^0_s$ mixing, as governed by the effective interaction after integrating out $Z'_{L,R}$ from the FCNCs, such as
\be \mathcal{L}^{\mathrm{eff}}_{\mathrm{FCNC}}\supset \fr{g^2(1+\beta^2 t^2_X)}{3 M^2}\left[|(V^*_{dL})_{32}(V_{dL})_{33}|^2 (\bar{s}_L\ga^\mu b_L)^2 +(L\rightarrow R)\right],\ee where $M$ is a typical mass for $Z'_{L,R}$, which is proportional to the new physics scales. From the data \cite{pdg}, we obtain
\be \fr{g^2(1+\beta^2 t^2_X)}{3 M^2}|(V^*_{dL})_{32}(V_{dL})_{33}|^2 <\fr{1}{(100\ \mathrm{TeV})^2}, \ee where $|(V^*_{dL})_{32}(V_{dL})_{33}|\simeq 3.9\times 10^{-2}$, assuming $V_{dL}=V_{\mathrm{CKM}}$. Thus, we get \be M>1.46\times \sqrt{1+\beta^2 t^2_X}\ \mathrm{TeV}\sim O(1)\ \mathrm{TeV}.\ee The $w$ scale is also in this order, in agreement with the bounds from the 750 GeV diphoton excess.                        

All the models discussed contain new gauge bosons that might be produced and then decay into jets or leptons, which are currently searched by the LHC. 
The bounds from the Drell-Yan processes may be evaded when the masses of $E$ and $\nu_R$ as well as the new neutral gauge bosons are large. We are not evaluating the new physics scales in detail, which is out of the scope of this paper. But, when $w$ is in $O(1)$ TeV, all the models are survival. But, when $w$ is in $O(10)$ TeV, only the least two models that possess the explicit left-right asymmetries may be alive for all the searches. Due to the mixings, the new gauge bosons of $SU(N)_R$ and $SU(N)_L/SU(2)_L$ can couple to the ordinary weak bosons such as $WW$, $ZZ$, and $WZ$, which provide a possible explanation of the ATLAS diboson excesses recently observed. Note that the constraints from the FCNCs, dijet and Drell-Yan processes as mentioned can change the results.   

To conclude, the $SU(3)_C\otimes SU(3)_L\otimes SU(3)_R\otimes U(1)_X$ model can provide the 750 GeV diphoton excess as a result of the spontaneous left-right symmetry breaking, where the new physics scale $w$ is in one TeV and the new electric charge $q$ is large, close to its bounds. The $SU(3)_C\otimes SU(2)_L\otimes SU(3)_R\otimes U(1)_X$ and $SU(3)_C\otimes SU(3)_L\otimes SU(2)_R\otimes U(1)_X$ models can explain the 750 diphoton excess naturally as a consequence of the explicit left-right asymmetry. All the other bounds can be evaded when the inert fields are interpreted instead, and the Yukawa coupling ratios (and the electric charge parameter for the second model) are free to float. Our theories show why there are only three generations of fermions observed in the nature.

\section*{Acknowledgments}
PVD would like to thank Dr. Avelino Vicente for the discussions on the 750 GeV diphoton excess in the 3-3-1 models.   
DTH is funded by Vietnam National Foundation for Science and Technology Development
(NAFOSTED) under grant number 103.01-2014.69.


\begin{thebibliography}{99}


\bibitem{MLR} J. C. Pati and A. Salam, Phys. Rev. D \textbf{10}, 275 (1974); R. N. Mohapatra and J. C. Pati,
Phys. Rev. D \textbf{11}, 566 (1975); R. N. Mohapatra and J. C. Pati, Phys. Rev. D \textbf{11}, 2558
(1975), G. Senjanovi\'c and R. N. Mohapatra, Phys. Rev. D \textbf{12}, 1502 (1975); G. Senjanovi\'c, Nucl.
Phys. B \textbf{153}, 334 (1979).

\bibitem{NLR} P. Minkowski, Phys. Lett. B \textbf{67}, 421 (1977);
 R. N. Mohapatra and G. Senjanovi\'c, Phys. Rev. Lett. \textbf{44}, 912 (1980);  R. N. Mohapatra and G. Senjanovi\'c,
Phys. Rev. D \textbf{23}, 165 (1981).

\bibitem{diphotontheor} See, for various proposals, 
K.~Harigaya and Y.~Nomura, arXiv:1512.04850 [hep-ph];
Y.~Mambrini, G.~Arcadi, and A.~Djouadi, arXiv:1512.04913 [hep-ph]; 
M.~Backovic, A.~Mariotti, and D.~Redigolo, arXiv:1512.04917 [hep-ph];
A.~Angelescu, A.~Djouadi, and G.~Moreau, arXiv:1512.04921 [hep-ph];
Y.~Nakai, R.~Sato, and K.~Tobioka, arXiv:1512.04924 [hep-ph];
S.~Knapen, T.~Melia, M.~Papucci, and K.~Zurek,   arXiv:1512.04928 [hep-ph];
D.~Buttazzo, A.~Greljo, and D.~Marzocca, arXiv:1512.04929 [hep-ph];
A.~Pilaftsis, arXiv:1512.04931 [hep-ph];
R.~Franceschini, G.~F. Giudice, J.~F. Kamenik, M.~McCullough, A.~Pomarol, R.~Rattazzi, M.~Redi, F.~Riva, A.~Strumia, and R.~Torre, arXiv:1512.04933 [hep-ph];
S.~Di~Chiara, L.~Marzola, and M.~Raidal, arXiv:1512.04939 [hep-ph];
T.~Higaki, K.~S. Jeong, N.~Kitajima, and F.~Takahashi, arXiv:1512.05295 [hep-ph];
S.~D. McDermott, P.~Meade, and H.~Ramani, arXiv:1512.05326 [hep-ph]; 
J.~Ellis, S.~A.~R. Ellis, J.~Quevillon, V.~Sanz, and T.~You, arXiv:1512.05327 [hep-ph]; 
M.~Low, A.~Tesi, and L.-T. Wang, arXiv:1512.05328 [hep-ph];
B.~Bellazzini, R.~Franceschini, F.~Sala, and J.~Serra, arXiv:1512.05330 [hep-ph];
C.~Petersson and R.~Torre, arXiv:1512.05333 [hep-ph];
R.~S. Gupta, S.~Jaeger, Y.~Kats, G.~Perez, and E.~Stamou, arXiv:1512.05332 [hep-ph];
E.~Molinaro, F.~Sannino, and N.~Vignaroli, arXiv:1512.05334 [hep-ph];
B.~Dutta, Y.~Gao, T.~Ghosh, I.~Gogoladze, and T.~Li, arXiv:1512.05439 [hep-ph];
Q.-H. Cao, Y.~Liu, K.-P. Xie, B.~Yan, and D.-M. Zhang, arXiv:1512.05542 [hep-ph];
S.~Matsuzaki and K.~Yamawaki, arXiv:1512.05564 [hep-ph];
A.~Kobakhidze, F.~Wang, L.~Wu, J.~M. Yang, and M.~Zhang, arXiv:1512.05585 [hep-ph];
R.~Martinez, F.~Ochoa, and C.~F. Sierra, arXiv:1512.05617 [hep-ph];
P.~Cox, A.~D. Medina, T.~S. Ray, and A.~Spray, arXiv:1512.05618 [hep-ph];
D.~Becirevic, E.~Bertuzzo, O.~Sumensari, and R.~Z. Funchal, arXiv:1512.05623 [hep-ph];
J.~M. No, V.~Sanz, and J.~Setford, arXiv:1512.05700 [hep-ph];
S.~V. Demidov and D.~S. Gorbunov, arXiv:1512.05723 [hep-ph];
W.~Chao, R.~Huo, and J.-H. Yu, arXiv:1512.05738 [hep-ph];
S.~Fichet, G.~von Gersdorff, and C.~Royon, arXiv:1512.05751 [hep-ph];
D.~Curtin and C.~B. Verhaaren, arXiv:1512.05753 [hep-ph];
L.~Bian, N.~Chen, D.~Liu, and J.~Shu, arXiv:1512.05759 [hep-ph];
J.~Chakrabortty, A.~Choudhury, P.~Ghosh, S.~Mondal, and T.~Srivastava, arXiv:1512.05767 [hep-ph];
A.~Ahmed, B.~M. Dillon, B.~Grzadkowski, J.~F. Gunion, and Y.~Jiang, arXiv:1512.05771 [hep-ph];
P.~Agrawal, J.~Fan, B.~Heidenreich, M.~Reece, and M.~Strassler, arXiv:1512.05775 [hep-ph];
C.~Csaki, J.~Hubisz, and J.~Terning, arXiv:1512.05776 [hep-ph];
A.~Falkowski, O.~Slone, and T.~Volansky,  arXiv:1512.05777 [hep-ph];
D.~Aloni, K.~Blum, A.~Dery, A.~Efrati, and Y.~Nir, arXiv:1512.05778 [hep-ph];
Y.~Bai, J.~Berger, and R.~Lu, arXiv:1512.05779 [hep-ph];
E.~Gabrielli, K.~Kannike, B.~Mele, M.~Raidal, C.~Spethmann, and H.~Veerme, arXiv:1512.05961 [hep-ph];
R.~Benbrik, C.-H. Chen, and T.~Nomura, arXiv:1512.06028 [hep-ph];
J.~S. Kim, J.~Reuter, K.~Rolbiecki, and R.~R. de~Austri, arXiv:1512.06083 [hep-ph];
A.~Alves, A.~G. Dias, and K.~Sinha, arXiv:1512.06091 [hep-ph];
E.~Megias, O.~Pujolas, and M.~Quiros,  arXiv:1512.06106 [hep-ph];
L.~M. Carpenter, R.~Colburn, and J.~Goodman, arXiv:1512.06107 [hep-ph];
J.~Bernon and C.~Smith, arXiv:1512.06113 [hep-ph];
W.~Chao, arXiv:1512.06297 [hep-ph];
M.~T. Arun and P.~Saha, arXiv:1512.06335 [hep-ph];
C.~Han, H.~M. Lee, M.~Park, and V.~Sanz,  arXiv:1512.06376 [hep-ph];
S.~Chang, arXiv:1512.06426 [hep-ph]; 
I.~Chakraborty and A.~Kundu, arXiv:1512.06508 [hep-ph];
R.~Ding, L.~Huang, T.~Li, and B.~Zhu, arXiv:1512.06560 [hep-ph];
H.~Han, S.~Wang, and S.~Zheng, arXiv:1512.06562 [hep-ph];
X.-F. Han and L.~Wang, arXiv:1512.06587 [hep-ph];
M.-x. Luo, K.~Wang, T.~Xu, L.~Zhang, and G.~Zhu, arXiv:1512.06670 [hep-ph];
J.~Chang, K.~Cheung, and C.-T. Lu, arXiv:1512.06671 [hep-ph];
D.~Bardhan, D.~Bhatia, A.~Chakraborty, U.~Maitra, S.~Raychaudhuri, and T.~Samui, arXiv:1512.06674 [hep-ph];
T.-F. Feng, X.-Q. Li, H.-B. Zhang, and S.-M. Zhao, arXiv:1512.06696 [hep-ph];
O.~Antipin, M.~Mojaza, and F.~Sannino, arXiv:1512.06708 [hep-ph];
F.~Wang, L.~Wu, J.~M. Yang, and M.~Zhang, arXiv:1512.06715 [hep-ph];
J.~Cao, C.~Han, L.~Shang, W.~Su, J.~M. Yang, and Y.~Zhang, arXiv:1512.06728 [hep-ph];
F.~P. Huang, C.~S. Li, Z.~L. Liu, and Y.~Wang, arXiv:1512.06732 [hep-ph];
W.~Liao and H.-q. Zheng, arXiv:1512.06741 [hep-ph];
J.~J. Heckman, arXiv:1512.06773 [hep-ph];
M.~Dhuria and G.~Goswami, arXiv:1512.06782 [hep-ph];
X.-J. Bi, Q.-F. Xiang, P.-F. Yin, and Z.-H. Yu, arXiv:1512.06787 [hep-ph];
J.~S. Kim, K.~Rolbiecki, and R.~R. de~Austri, arXiv:1512.06797 [hep-ph];
L.~Berthier, J.~M. Cline, W.~Shepherd, and M.~Trott, arXiv:1512.06799 [hep-ph];
W.~S. Cho, D.~Kim, K.~Kong, S.~H. Lim, K.~T. Matchev, J.-C. Park, and M.~Park, arXiv:1512.06824 [hep-ph];
J.~M. Cline and Z.~Liu, arXiv:1512.06827 [hep-ph];
M.~Bauer and M.~Neubert,  arXiv:1512.06828 [hep-ph];
M.~Chala, M.~Duerr, F.~Kahlhoefer, and K.~Schmidt-Hoberg, arXiv:1512.06833 [hep-ph];
K.~Kulkarni, arXiv:1512.06836 [hep-ph];
D.~Barducci, A.~Goudelis, S.~Kulkarni, and D.~Sengupta, arXiv:1512.06842 [hep-ph];
S. M. Boucenna, S. Morisi, and A. Vicente, arXiv:1512.06878 [hep-ph];
C.~W. Murphy,  arXiv:1512.06976 [hep-ph];
A.~E.~C. Hernandez and I.~Nisandzic, arXiv:1512.07165 [hep-ph];
U.~K. Dey, S.~Mohanty, and G.~Tomar, arXiv:1512.07212 [hep-ph];
G.~M. Pelaggi, A.~Strumia, and E.~Vigiani, arXiv:1512.07225 [hep-ph];
J.~de~Blas, J.~Santiago, and R.~Vega-Morales, arXiv:1512.07229 [hep-ph];
A.~Belyaev, G.~Cacciapaglia, H.~Cai, T.~Flacke, A.~Parolini, and H.~Serodio, arXiv:1512.07242 [hep-ph];
P.~S.~B. Dev and D.~Teresi, arXiv:1512.07243 [hep-ph];
W.-C. Huang, Y.-L.~S. Tsai, and T.-C. Yuan, arXiv:1512.07268 [hep-ph];
S.~Moretti and K.~Yagyu, arXiv:1512.07462 [hep-ph];
K.~M. Patel and P.~Sharma, arXiv:1512.07468 [hep-ph];
M.~Badziak,  arXiv:1512.07497 [hep-ph];
S.~Chakraborty, A.~Chakraborty, and S.~Raychaudhuri, arXiv:1512.07527 [hep-ph];
Q.-H. Cao, S.-L. Chen, and P.-H. Gu, arXiv:1512.07541 [hep-ph];
W.~Altmannshofer, J.~Galloway, S.~Gori, A.~L. Kagan, A.~Martin, and J.~Zupan,arXiv:1512.07616 [hep-ph];
M.~Cveti, J.~Halverson, and P.~Langacker, arXiv:1512.07622 [hep-ph];
J.~Gu and Z.~Liu, arXiv:1512.07624 [hep-ph];
B.~C. Allanach, P.~S.~B. Dev, S.~A. Renner, and K.~Sakurai, arXiv:1512.07645 [hep-ph];
H.~Davoudiasl and C.~Zhang, arXiv:1512.07672 [hep-ph];
N.~Craig, P.~Draper, C.~Kilic, and S.~Thomas, arXiv:1512.07733 [hep-ph];
K.~Das and S.~K. Rai, arXiv:1512.07789 [hep-ph];
K.~Cheung, P.~Ko, J.~S. Lee, J.~Park, and P.-Y. Tseng, arXiv:1512.07853 [hep-ph];
J.~Liu, X.-P. Wang, and W.~Xue, arXiv:1512.07885 [hep-ph];
J.~Zhang and S.~Zhou, arXiv:1512.07889 [hep-ph];
J.~A. Casas, J.~R. Espinosa, and J.~M. Moreno, arXiv:1512.07895 [hep-ph];
L.~J. Hall, K.~Harigaya, and Y.~Nomura,  arXiv:1512.07904 [hep-ph]; 
H.~Han, S.~Wang, and S.~Zheng, arXiv:1512.07992 [hep-ph];
J.-C. Park and S.~C. Park, arXiv:1512.08117 [hep-ph];
A.~Salvio and A.~Mazumdar, arXiv:1512.08184 [hep-ph];
D.~Chway, R.~Dermisek, T.~H. Jung, and H.~D. Kim, arXiv:1512.08221 [hep-ph];
G.~Li, Y.-n. Mao, Y.-L. Tang, C.~Zhang, Y.~Zhou, and S.-h. Zhu, arXiv:1512.08255 [hep-ph];
M.~Son and A.~Urbano, arXiv:1512.08307 [hep-ph];
Y.-L. Tang and S.-h. Zhu, arXiv:1512.08323 [hep-ph];
H.~An, C.~Cheung, and Y.~Zhang, arXiv:1512.08378 [hep-ph];
J.~Cao, F.~Wang, and Y.~Zhang, arXiv:1512.08392 [hep-ph];
F.~Wang, W.~Wang, L.~Wu, J.~M. Yang, and M.~Zhang, arXiv:1512.08434 [hep-ph];
C.~Cai, Z.-H. Yu, and H.-H. Zhang, arXiv:1512.08440 [hep-ph];
Q.-H. Cao, Y.~Liu, K.-P. Xie, B.~Yan, and D.-M. Zhang, arXiv:1512.08441 [hep-ph];
J.~E. Kim, arXiv:1512.08467 [hep-ph];
J.~Gao, H.~Zhang, and H.~X. Zhu, arXiv:1512.08478 [hep-ph];
W.~Chao, arXiv:1512.08484 [hep-ph];
X.-J. Bi, R.~Ding, Y.~Fan, L.~Huang, C.~Li, T.~Li, S.~Raza, X.-C. Wang, and B.~Zhu, arXiv:1512.08497 [hep-ph];
F.~Goertz, J.~F. Kamenik, A.~Katz, and M.~Nardecchia, arXiv:1512.08500 [hep-ph];
L.~A. Anchordoqui, I.~Antoniadis, H.~Goldberg, X.~Huang, D.~Lust, and T.~R. Taylor, arXiv:1512.08502 [hep-ph];
P.~S.~B. Dev, R.~N. Mohapatra, and Y.~Zhang, arXiv:1512.08507 [hep-ph];
N.~Bizot, S.~Davidson, M.~Frigerio, and J.~L. Kneur, arXiv:1512.08508 [hep-ph];
L.~E. Ibanez and V.~Martin-Lozano, arXiv:1512.08777 [hep-ph];
C.-W. Chiang, M.~Ibe, and T.~T. Yanagida, arXiv:1512.08895 [hep-ph];
S.~K. Kang and J.~Song, arXiv:1512.08963 [hep-ph];
Y.~Hamada, T.~Noumi, S.~Sun, and G.~Shiu, arXiv:1512.08984 [hep-ph];
X.-J. Huang, W.-H. Zhang, and Y.-F. Zhou, arXiv:1512.08992 [hep-ph];
S.~Kanemura, K.~Nishiwaki, H.~Okada, Y.~Orikasa, S.~C. Park, and R.~Watanabe, arXiv:1512.09048 [hep-ph];
S.~Kanemura, N.~Machida, S.~Odori, and T.~Shindou, arXiv:1512.09053 [hep-ph];
I.~Low and J.~Lykken, arXiv:1512.09089 [hep-ph];
A.~E.~C. Hernandez, arXiv:1512.09092 [hep-ph];
Y.~Jiang, Y.-Y. Li, and T.~Liu, arXiv:1512.09127 [hep-ph];
K.~Kaneta, S.~Kang, and H.-S. Lee, arXiv:1512.09129 [hep-ph];
L.~Marzola, A.~Racioppi, M.~Raidal, F.~R. Urban, and H.~Veermae, arXiv:1512.09136 [hep-ph];
A.~Dasgupta, M.~Mitra, and D.~Borah, arXiv:1512.09202 [hep-ph];
S.~Jung, J.~Song, and Y.~W. Yoon, arXiv:1601.00006 [hep-ph];
C.~T. Potter, arXiv:1601.00240 [hep-ph];
E.~Palti, arXiv:1601.00285 [hep-ph];
T.~Nomura and H.~Okada, arXiv:1601.00386 [hep-ph];
X.-F. Han, L.~Wang, L.~Wu, J.~M. Yang, and M.~Zhang, arXiv:1601.00534 [hep-ph];
P.~Ko, Y.~Omura, and C.~Yu, arXiv:1601.00586 [hep-ph];
K.~Ghorbani and H.~Ghorbani, arXiv:1601.00602 [hep-ph];
D.~Palle, arXiv:1601.00618 [hep-ph];
U.~Danielsson, R.~Enberg, G.~Ingelman, and T.~Mandal,  arXiv:1601.00624 [hep-ph];
W.~Chao, arXiv:1601.00633 [hep-ph];
C.~Csaki, J.~Hubisz, S.~Lombardo, and J.~Terning, arXiv:1601.00638 [hep-ph];
A.~Karozas, S.~F. King, G.~K. Leontaris, and A.~K. Meadowcroft, arXiv:1601.00640 [hep-ph];  
A.~E.~C. Hernandez, I.~d.~M. Varzielas, and E.~Schumacher, arXiv:1601.00661 [hep-ph];
T.~Modak, S.~Sadhukhan, and R.~Srivastava, arXiv:1601.00836 [hep-ph];
B.~Dutta, Y.~Gao, T.~Ghosh, I.~Gogoladze, T.~Li, Q.~Shafi, and J.~W. Walker, arXiv:1601.00866 [hep-ph];
F.~F. Deppisch, C.~Hati, S.~Patra, P.~Pritimita, and U.~Sarkar, arXiv:1601.00952 [hep-ph];
H.~Ito, T.~Moroi, and Y.~Takaesu, arXiv:1601.01144 [hep-ph];
H.~Zhang, arXiv:1601.01355 [hep-ph];
A.~Berlin, arXiv:1601.01381 [hep-ph];
S.~Bhattacharya, S.~Patra, N.~Sahoo, and N.~Sahu, arXiv:1601.01569 [hep-ph];
F.~D'Eramo, J.~de~Vries, and P.~Panci, arXiv:1601.01571 [hep-ph];
I.~Sahin,  arXiv:1601.01676 [hep-ph];
S.~Fichet, G.~von Gersdorff, and C.~Royon, arXiv:1601.01712 [hep-ph];
D.~Borah, S.~Patra, and S.~Sahoo, arXiv:1601.01828 [hep-ph]; 
D.~Stolarski and R.~Vega-Morales, arXiv:1601.02004 [hep-ph];
C.~Hati, arXiv:1601.02457 [hep-ph];
M.~T. Arun and D.~Choudhury, arXiv:1601.02321 [hep-ph];
M.~Fabbrichesi and A.~Urbano, arXiv:1601.02447 [hep-ph];
P.~Ko and T.~Nomura, arXiv:1601.02490 [hep-ph];
J.~Cao, L.~Shang, W.~Su, Y.~Zhang, and J.~Zhu, arXiv:1601.02570 [hep-ph];
J.-H. Yu,  arXiv:1601.02609 [hep-ph];
R.~Ding, Z.-L. Han, Y.~Liao, and X.-D. Ma,  arXiv:1601.02714 [hep-ph];
S.~Alexander and L.~Smolin, arXiv:1601.03091 [hep-ph]; 
J.~H. Davis, M.~Fairbairn, J.~Heal, and P.~Tunney, arXiv:1601.03153 [hep-ph];
L.~V. Laperashvili, H.~B. Nielsen, and C.~R. Das, arXiv:1601.03231 [hep-ph];
I.~Dorsner, S.~Fajfer, and N.~Kosnik, arXiv:1601.03267 [hep-ph];
A.~E. Faraggi and J.~Rizos, arXiv:1601.03604 [hep-ph];
A.~Djouadi, J.~Ellis, R.~Godbole, and J.~Quevillon, arXiv:1601.03696 [hep-ph];
L.~A. Harland-Lang, V.~A. Khoze, and M.~G. Ryskin, arXiv:1601.03772 [hep-ph];
T.~Nomura and H.~Okada, arXiv:1601.04516 [hep-ph];
W.~Chao, arXiv:1601.04678 [hep-ph];
M.~R. Buckley, arXiv:1601.04751 [hep-ph];
A.~Ghoshal, arXiv:1601.04291 [hep-ph];
X.-F. Han, L.~Wang, and J.~M. Yang, arXiv:1601.04954 [hep-ph];
H.~Okada and K.~Yagyu, arXiv:1601.05038 [hep-ph];
D.~B. Franzosi and M.~T. Frandsen, arXiv:1601.05357 [hep-ph];
A.~Martini, K.~Mawatari, and D.~Sengupta, arXiv:1601.05729 [hep-ph];
Q.-H. Cao, Y.-Q. Gong, X.~Wang, B.~Yan, and L.~L. Yang, arXiv:1601.06374 [hep-ph];
C.-W. Chiang and A.-L. Kuo, arXiv:1601.06394 [hep-ph];
U.~Aydemir and T.~Mandal, arXiv:1601.06761 [hep-ph];
S.~Abel and V.~V. Khoze, arXiv:1601.07167 [hep-ph];
L.~A. Harland-Lang, V.~A. Khoze, and M.~G. Ryskin,  arXiv:1601.07187 [hep-ph];
S.~F. King and R.~Nevzorov, arXiv:1601.07242 [hep-ph];
T.~Nomura and H.~Okada,  arXiv:1601.07339 [hep-ph];
C.-Q. Geng and D.~Huang, arXiv:1601.07385 [hep-ph];
B.~J. Kavanagh, arXiv:1601.07330 [hep-ph];
J.~Kawamura and Y.~Omura, arXiv:1601.07396 [hep-ph];
E.~Bertuzzo, P.~A.~N. Machado, and M.~Taoso, arXiv:1601.07508 [hep-ph];
I.~Ben-Dayan and R.~Brustein,arXiv:1601.07564 [hep-ph];
A.~D. Martin and M.~G. Ryskin, arXiv:1601.07774 [hep-ph];
A.~Hektor and L.~Marzola, arXiv:1602.00004 [hep-ph];
N.~D. Barrie, A.~Kobakhidze, M.~Talia, and L.~Wu, arXiv:1602.00475 [hep-ph];
L.~Aparicio, A.~Azatov, E.~Hardy, and A.~Romanino, arXiv:1602.00949 [hep-ph];
R.~Ding, Y.~Fan, L.~Huang, C.~Li, T.~Li, S.~Raza, and B.~Zhu, arXiv:1602.00977 [hep-ph];
K.~Harigaya and Y.~Nomura, arXiv:1602.01092 [hep-ph];
T.~Li, J.~A. Maxin, V.~E. Mayes, and D.~V. Nanopoulos, arXiv:1602.01377 [hep-ph];
A.~Salvio, F.~Staub, A.~Strumia, and A.~Urbano, arXiv:1602.01460 [hep-ph];
S.-F. Ge, H.-J. He, J.~Ren, and Z.-Z. Xianyu, arXiv:1602.01801 [hep-ph];
S.~I. Godunov, A.~N. Rozanov, M.~I. Vysotsky, and E.~V. Zhemchugov, arXiv:1602.02380 [hep-ph];
S.~B. Giddings and H.~Zhang, arXiv:1602.02793 [hep-ph];
U.~Ellwanger and C.~Hugonie, arXiv:1602.03344 [hep-ph];
P.~Draper and D.~McKeen, arXiv:1602.03604 [hep-ph];
C.~Arbelaez, A.~E.~C. Hernandez, S.~Kovalenko, and I.~Schmidt, arXiv:1602.03607 [hep-ph];
K.~J. Bae, M.~Endo, K.~Hamaguchi, and T.~Moroi, arXiv:1602.03653 [hep-ph];
C.~Gross, O.~Lebedev, and J.~M. No, arXiv:1602.03877 [hep-ph];
Y.~Hamada, H.~Kawai, K.~Kawana, and K.~Tsumura, arXiv:1602.04170 [hep-ph];
C.~Han, T.~T. Yanagida, and N.~Yokozaki, arXiv:1602.04204 [hep-ph];
B.~Dasgupta, J.~Kopp, and P.~Schwaller, arXiv:1602.04692 [hep-ph];
C.~Frugiuele, E.~Fuchs, G.~Perez, and M.~Schlaffer, arXiv:1602.04822 [hep-ph]
C.~Delaunay and Y.~Soreq, arXiv:1602.04838 [hep-ph];
Y.-J. Zhang, B.-B. Zhou, and J.-J. Sun, arXiv:1602.05539 [hep-ph];
F. Staub, P. Athron, L. Basso, M. D. Goodsell, D. Harries, M. E. Krauss, K. Nickel, T. Opferkuch, L. Ubaldi, A. Vicente, A. Voigt, arXiv:1602.05581 [hep-ph];
S. Baek and J.-h. Park, arXiv:1602.05588 [hep-ph];
G. Lazarides and Q. Shafi, arXiv:1602.07866 [hep-ph];
C. Bonilla, M. Nebot, R. Srivastava, and J. W. F. Valle, arXiv:1602.08092 [hep-ph];
T. Li, J. A. Maxin, V. E. Mayes, and D. V. Nanopoulos, arXiv:1602.09099 [hep-ph];
J. Bernon, A. Goudelis, S. Kraml, K. Mawatari, and D. Sengupta, arXiv:1603.03421 [hep-ph]. 

\bibitem{diphotonexp} The ATLAS Collaboration, ATLAS-CONF-2015-081; The CMS Collaboration, CMS-PAS-EXO-15-004. 

\bibitem{331m} F. Pisano and V. Pleitez, Phys. Rev.  D {\bf 46}, 410 (1992); P. H. Frampton, Phys. Rev. Lett. {\bf 69}, 2889 (1992); R. Foot, O. F. Hernandez, F. Pisano, and V. Pleitez, Phys. Rev. D {\bf 47}, 4158 (1993).

\bibitem{331r} M. Singer, J. W. F. Valle, and J. Schechter, Phys. Rev. D {\bf 22}, 738 (1980); J. C. Montero, F. Pisano, and V. Pleitez, Phys. Rev. D {\bf 47}, 2918 (1993); R. Foot, H. N. Long, and Tuan A. Tran, Phys. Rev. D {\bf 50}, 34 (1994); P. V. Dong, H. N. Long, D. T. Nhung, and D. V. Soa, Phys. Rev. D {\bf 73}, 035004 (2006); S. M. Boucenna, J. W. F. Valle, and A. Vicente, Phys. Rev. D {\bf 92}, 053001 (2015); J. W. F. Valle and C. A. Vaquera-Araujo, Phys. Lett. B {\bf 755}, 363 (2016).   

\bibitem{dibosonexp} G. Aad {\it et al.} (ATLAS Collaboration), JHEP {\bf 1512}, 055 (2015).

\bibitem{dps} Alex G. Dias, C. A. de S. Pires, and P. S. Rodrigues da Silva, Phys. Rev. D {\bf 82}, 035013 (2010); Cesar P. Ferreira, Marcelo M. Guzzo, and Pedro C. de Holanda, arXiv:1509.02977 [hep-ph]. 

\bibitem{pdg} K. A. Olive {\it et al.} (Particle Data Group), Chin. Phys. C {\bf 38}, 090001 (2014), and partial updates at http://pdg.lbl.gov.

\bibitem{trinification} Y. Achiman and B. Stech, in {\it New Phenomena in Lepton-Hadron Physics}, edited by D. E. C. Fries and J. Wess (Plenum, New York, 1979), p. 303; A. de Rujula, H. Georgi, and S. L. Glashow, in {\it Fifth Workshop on Grand Unification}, edited by K. Kang, H. Fried, and P. Frampton (World Scientific, Singapore, 1984), p. 88; J. Hetzel and B. Stech, Phys. Rev. D {\bf 91}, 055026 (2015); G.M. Pelaggi, A. Strumia, and S. Vignali, JHEP {\bf 1508},  130 (2015). 

\bibitem{strumia} R. Franceschini {\it et al.} in \cite{diphotontheor}.  

\bibitem{dn} P. V. Dong and N. T. K. Ngan, arXiv:1512.09073 [hep-ph].

\bibitem{beta331} P. V. Dong, Phys. Rev. D {\bf 92}, 055026 (2015); See also, R. A. Diaz, R. Martinez, and F. Ochoa, Phys. Rev. D {\bf 72}, 035018 (2005).

\end{thebibliography}
 \end{document}